\documentclass[%
 aip,
 amsmath,amssymb,
 reprint,%
]{revtex4-1}

\usepackage{graphicx}
\usepackage{dcolumn}
\usepackage{bm}
\usepackage{color}
\usepackage[utf8]{inputenc}
\usepackage[T1]{fontenc}
\usepackage{mathptmx}
\usepackage{etoolbox}
\usepackage{hyperref}
\hypersetup{colorlinks=true, linkcolor=blue,citecolor=blue,urlcolor=blue}
\usepackage{diagbox}
\usepackage{booktabs}
\usepackage[table,xcdraw]{xcolor}
\usepackage{colortbl}
\definecolor{tabcolor}{rgb}{.105,.410,.113}

\makeatletter
\def\@email#1#2{%
 \endgroup
 \patchcmd{\titleblock@produce}
  {\frontmatter@RRAPformat}
  {\frontmatter@RRAPformat{\produce@RRAP{*#1\href{mailto:#2}{#2}}}\frontmatter@RRAPformat}
  {}{}
}%
\makeatother
\begin{document}

\preprint{AIP/123-QED}

\title[Sample title]{Evolution of cooperation with Q-learning: the impact of information perception}
\author{Guozhong Zheng}
 \affiliation{School of Physics and Information Technology, Shaanxi Normal University, Xi'an 710061, P. R. China}
\author{Zhenwei Ding}
\affiliation{School of Physics, Ningxia University, Yinchuan 750021, P. R. China}
\affiliation{School of Xinjiang Institute of Engineering Control Engineering College, Xinjiang Institute of Engineering, Ürümqi 830023, P. R. China}
\author{Jiqiang Zhang}
\affiliation{School of Physics, Ningxia University, Yinchuan 750021, P. R. China}
\author{Shengfeng Deng}
\affiliation{School of Physics and Information Technology, Shaanxi Normal University, Xi'an 710061, P. R. China}
\author{Weiran Cai}
\affiliation{School of Computer Science, Soochow University, Suzhou 215006, P. R. China}
\author{Li Chen}
\homepage[Email address: ]{chenl@snnu.edu.cn}
\affiliation{School of Physics and Information Technology, Shaanxi Normal University, Xi'an 710061, P. R. China}

\date{\today}

\begin{abstract}
The inherent complexity of human beings manifests in a remarkable diversity of responses to intricate environments, enabling us to approach problems from varied perspectives. However, in the study of cooperation, existing research within the reinforcement learning framework often assumes that individuals have access to identical information when making decisions, which contrasts with the reality that individuals frequently perceive information differently. In this study, we employ the Q-learning algorithm to explore the impact of information perception on the evolution of cooperation in a two-person Prisoner's Dilemma game. We demonstrate that the evolutionary processes differ significantly across three distinct information perception scenarios, highlighting the critical role of information structure in the emergence of cooperation. Notably, the asymmetric information scenario reveals a complex dynamical process, including the emergence, breakdown, and reconstruction of cooperation, mirroring psychological shifts observed in human behavior. Our findings underscore the importance of information structure in fostering cooperation, offering new insights into the establishment of stable cooperative relationships among humans.
\end{abstract}

\maketitle

\begin{quotation}
In the real world, our perceptions of information are shaped by a variety of factors, leading to diverse responses to environmental stimuli and underscoring the importance of perceptual differences in decision-making processes. To explore how these differences influence the evolution of cooperation, we develop a simplified two-player Prisoner's Dilemma model using the Q-learning algorithm. By analyzing three distinct information perception scenarios, we observe significantly different evolutionary processes, with the asymmetric information scenario exhibiting particularly complex dynamics in the emergence and stability of cooperation. These findings emphasize the critical role of information structure in shaping cooperative behaviors and provide new insights into the complexities of human decision-making.
\end{quotation}

\section{\label{sec:introduction}Introduction}

Cooperation is fundamental to the survival, development, and reproduction of humans and other species, playing a crucial role in improving collective efficiency and benefits~\cite{axelrod1981evolution,Maynard1995The,Marko2022Social}. However, its complexity and subtlety often lead to non-cooperation, manifesting in issues like global warming, overfishing, and conflicts, which can have catastrophic consequences. Understanding how cooperation emerges and under what conditions it breaks down remains a central challenge~\cite{Elizabeth2005How}. Evolutionary game theory~\cite{Szabo2007Evolutionary,Perc2010Coevolutionary}, particularly through models like the prisoner's dilemma (PD) game~\cite{rapoport1965prisoner}, has been instrumental in studying cooperation. The PD game illustrates the difficulty of maintaining cooperation despite its collective benefits, as individuals tend to prioritize self-interest and defect. Identifying mechanisms that overcome this dilemma to promote cooperation is therefore essential.

Several mechanisms for the emergence of cooperation have been proposed in the past decades~\cite{Nowak2006Five,PERC2017Statistical}, including direct~\cite{trivers1971evolution} and indirect reciprocity~\cite{nowak1998evolution}, kin and group selection~\cite{Queller1964Group}, punishment and reward~\cite{Sigmund2001Reward}, network~\cite{nowak1992evolutionary, Szab1998Evolutionary, Wang2013Interdependent} and dynamical reciprocity~\cite{Liang2022dynamical}, social diversity~\cite{Perc2008Social, Santos2008Social, Liang2021Social}, reputation~\cite{Xia2023Reputation}, and behavioral multimodality~\cite{Ma2023emergence} etc.
Note that these game-theoretic studies typically employ imitation learning~\cite{Roca2009Evolutionary}, such as the Moran rule~\cite{Knight2018Evolution}, Fermi-function-based update rule~\cite{Szab1998Evolutionary, Szabo2005Phase}, and follow-the-best rule~\cite{Nowak1993strategy} et al. The idea behind is that individuals are more likely to imitate strategies of neighbors who are better off, which can be regarded as a simplified form of social learning~\cite{Bandura1977social}.

Reinforcement learning (RL)~\cite{Sutton2018reinforcement} as an alternative paradigm provides a fundamentally different perspective to study the evolution of cooperation~\cite{Wang2024Mathematics}. In RL, players score different actions within different states, and by repeatedly interacting with the environment they are able to make decisions by balancing the past experience, the present reward, and the expected earnings in the future. Despite its great potential~\cite{Lee2008Game,Silver2016Mastering,Subramanian2022Reinforcement}, RL as a distinct learning paradigm from imitation learning has been largely overlooked. Recently, researchers have started to apply reinforcement learning to evolutionary game theory to help understand the evolution of social behaviors, such as cooperation~\cite{Tanabe2012Evolution,Fan2022Incorporating,Shi2022analysis,Zhang2020Oscillatory,Wang2022Levy,Wang2023Synergistic,He2022migration,Ding2023Emergence,Geng2022Reinforcement,Zhang2024emergence,Zheng2024Evolution,mintz2025evolutionary}, trust~\cite{Zheng2024decoding}, fairness~\cite{zheng2024decodingfair}, collective motion~\cite{Incera2020Development,Wang2023Modeling}, and resource allocation~\cite{Andrecut2001q,Zhang2019reinforcement}.

This growing body of work highlights the versatility of RL in understanding complex social dynamics and its potential to uncover new insights into the mechanisms driving cooperative and collective behaviors. For instance, Zhang et al. demonstrated that explosive cooperation manifests as periodic oscillations in snowdrift games using RL~\cite{Zhang2020Oscillatory}. Wang et al. found that L\'{e}vy noise enhances cooperation through RL, accounting for real-world uncertainties~\cite{Wang2022Levy}. Later, they integrated an adaptive reward mechanism into the public goods game, showing a significant increase in cooperation levels~\cite{Wang2023Synergistic}. He et al. extended the PD game to mobile populations, revealing that adaptive migration strengthens network reciprocity and promotes cooperation in dense populations~\cite{He2022migration}. In two-player scenarios, Ding et al. showed that coordinated optimal policies emerge from strong memory and long-term expectations, with agents adopting a "win-stay, lose-shift" strategy to sustain high cooperation~\cite{Ding2023Emergence}. Additionally, studies suggest that RL can catalyze cooperation when combined with other updating rules~\cite{Geng2022Reinforcement, Han2022hybrid, Sheng2024catalytic}. However, these works assume symmetric information perception, where individuals access the same type of information, such as their own actions~\cite{Zhang2020Oscillatory,Wang2022Levy,Wang2023Synergistic}, neighbors' actions~\cite{He2022migration}, or both~\cite{Ding2023Emergence,Geng2022Reinforcement}.

Yet, numerous real-world observations indicate that information perception is often asymmetric, shaped by factors like age, experience, culture, social status, and personal beliefs~\cite{dawkins1989selfish,Molinas1998The}, as well as indirect influences such as economic, social, and political environments~\cite{Henrich2004Foundations}. This diversity leads individuals to focus on different aspects of available information~\cite{John1996Fairness,Allison1978Measures,Feltovich2000Reinforcement,McAvoy2015Asymmetric}, raising the question of \emph{how such variations in information perception impact cooperation}. 
While some studies highlight the role of information richness in cooperation~\cite{Jia2021Local,Yang2024Interaction,Song2022Reinforcement}, they often rely on network structures and neighbor payoff information, leaving the more fundamental pairwise interactions and action-based information unexplored.

In this work, we adopt a fresh perspective by distinguishing between information sources within the RL framework, focusing on action information rather than payoffs. Using the Q-learning algorithm~\cite{Watkins1989learning,Watkins1992Q}, we systematically investigate cooperation evolution under symmetric and asymmetric information settings in two-player PD games~\cite{Doebeli2005Models}. We identify distinct mechanisms across three information perception scenarios, revealing rich dynamical behaviors in the asymmetric case, including cooperation emergence, breakdown, and reestablishment. Notably, the asymmetric scenario achieves the highest cooperation preference in the shortest time, underscoring the critical role of information structure in shaping cooperative dynamics.

This paper is organized as follows:
we introduce our Q-learning model with three different information schemes in Sec.~\ref{sec:model}.
In Sec.~\ref{sec:results}, we present the results.
In Sec.~\ref{sec:analysis}, we provide a mechanistic analysis.
In Sec.~\ref{sec:compare}, the evolution processes for both symmetric and asymmetric information scenarios are compared.
Finally, we conclude our work together with discussions in Sec.~\ref{sec:discussion}.

\section{Model}\label{sec:model}

\begin{figure}[t]
\centering
\includegraphics[width=0.9\linewidth]{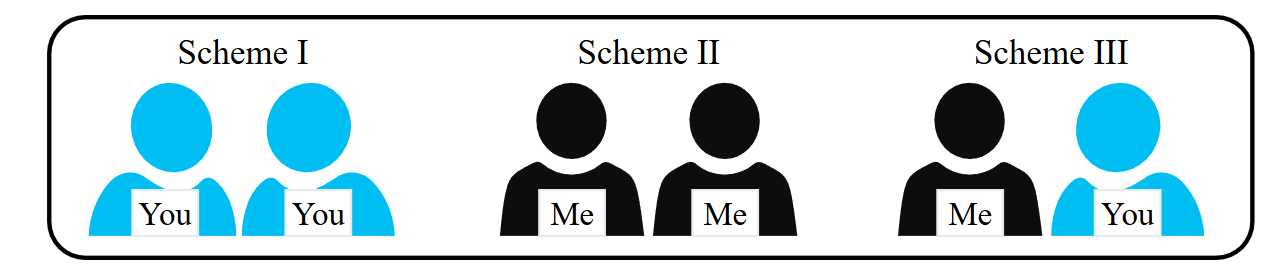}
\caption{\textbf{Three information schemes for playing a pairwise game.}
Scheme I -- ``You + You'' and  Scheme II -- ``Me + Me'' are both symmetric, but Scheme III -- ``You + Me'' is asymmetric and both consider the action information of the blue player labeled with ``Me''.}
\label{fig:ThreeSchemes}
\end{figure}

We consider the two-player scenario where they play the prisoner's dilemma game (PD), each having two options: cooperation (C) or defection (D). Mutual cooperation brings each a reward $R$, while mutual defection leads to a punishment $P$ for each. The mixed encounter scenario brings the cooperator the sucker's payoff $S$ and the defector the temptation $T$. The payoffs need to satisfy $T \textgreater R \textgreater P \textgreater S$, and $T+S < 2R$ for collective concern. The payoff matrix is summarized as follows:
\begin{equation}
\boldsymbol{\Pi}=\left(\begin{array}{ll}
\Pi_{C C} & \Pi_{C D} \\
\Pi_{D C} & \Pi_{D D}
\end{array}\right)=\left(\begin{array}{cc}
R & S \\
T & P
\end{array}\right),
\label{eq:payoff}
\end{equation}
where $R=1.0$, $S=-b$, $T=1+b$, and $P=0$ are adopted in our study, corresponding to a strong version of PD~\cite{Axelrod1981The}. $b>0$ is the dilemma strength, a larger value of which means less likely for cooperation to survive. A more general understanding of dilemma strength in symmetric 2×2 games can be found in refs.~\cite{Tanimoto2007Relationship,Hiromu2018Scaling}.

\begin{figure*}[bpth]
\centering
\includegraphics[width=0.8\linewidth]{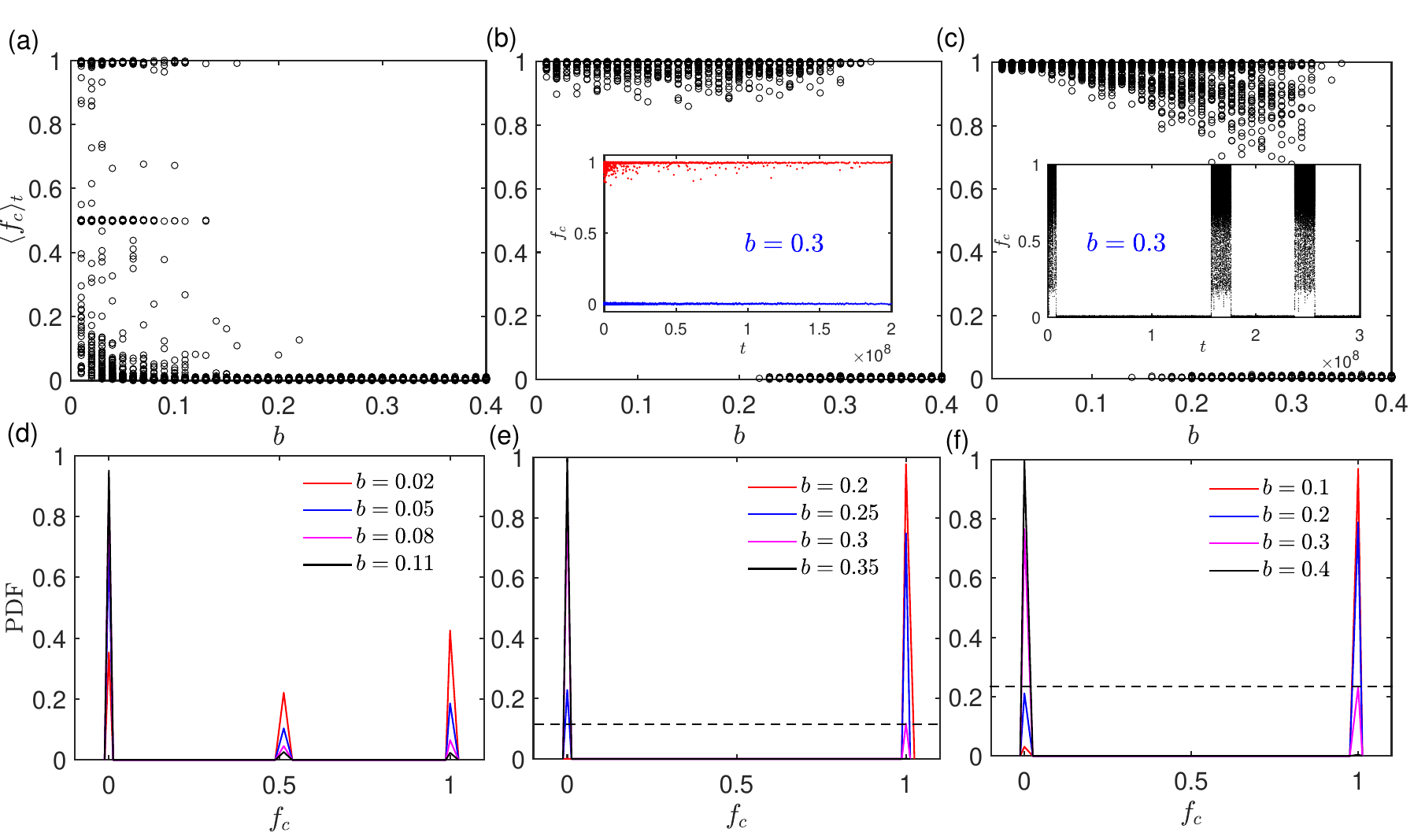}
\caption{\textbf{The dependence of cooperation preference on the dilemma strength within the three schemes.}
(a-c) The time-averaged cooperation preference $\left \langle f_c \right \rangle_t$ versus the dilemma strength $b$, respectively for Scheme I, II, III.
While no clear dependence is observed in Scheme I, the dependence shows a discontinuous transition of cooperation preference in Scheme II and III.
The two insets show typical time series of $f_c$ for $b=0.3$ in the corresponding scheme; the red and blue lines represent the results of evolution from two different initial conditions in (b). This means that once the system evolves into mutual cooperation or mutual defection, no change is expected. But persist
 state switches between the two solutions are always observed in (c).
(d-f) The corresponding probability density function (PDF) curve of $f_c$, respectively, for Scheme I-III, where trimodal distribution is seen for Scheme I, and bimodal distributions are for the other two schemes.
The dashed lines in (e, f) indicate the peak value of $f_c=1$ where $b=0.3$, where a higher value is observed in Scheme III than Scheme II.
Each data is averaged 500 times after a transient of $3\times10^8$ rounds in (a)-(c).
Other parameters: $\epsilon$ = 0.01, $\alpha$ = 0.1, $\gamma$ = 0.9.
}
\label{fig:phasetransition}
\end{figure*}

In our work, players adopt the Q-learning algorithm~\cite{Watkins1992Q}, where their decision-making is guided by a two-dimensional table termed as Q-table.
 The Q-table in our study is as follows:

\begin{table}[h]
\begin{tabular}{c|cc}
\arrayrulecolor{tabcolor}\toprule [1.4pt]
\hline
\diagbox{State}{Action}& C ($a_{1}$) & D ($a_{2}$) \\
\hline
C  ($s_{1}$) & $Q_{s_{1},a_{1}}$ & $Q_{s_{1},a_{2}}$  \\
D ($s_{2}$) & $Q_{s_{2},a_{1}}$ & $Q_{s_{2},a_{2}}$  \\
\hline
\bottomrule[1.4pt]
\end{tabular}
\label{tab:QtableA}
\end{table}
The state set $\mathbb{S}=\{$C$, $D$\}$ and the action set $\mathbb{A}=\{$C$, $D$\}$ are formally identical and simple.
The items in the table are Q-value $Q_{s,a}$, which scores the value of the action $a\in \mathbb{A}$ within the given state $s\in \mathbb{S}$. With a larger value of $Q_{s,a}>Q_{s,\hat{a}}$, the action $a$ is more preferred than $\hat{a}$ within the state $s$.
While the action information available to players is definite, the set of states $\mathbb{S}$ reflects the information about the environment that individuals perceive. Different players could have different perceived information (i.e., the state set $\mathbb{S}$) which they may find useful.

Specifically, we consider three different information schemes.
(I) Both players are informed of the opponent's action;
(II) Both players consider one's own action information;
(III) One player considers the opponent's action information, while the other considers one's own action information in the last round. Obviously, in either Scheme I or II, the information used is structurally symmetric for the two players, but this is not the case in Scheme III, where they both concern the action of one player, and is thus asymmetric. The illustration of the three schemes is shown in Fig.~\ref{fig:ThreeSchemes}.

The evolution of the two-player system follows a synchronous updating procedure. At the beginning, each player is randomly assigned an initial strategy C or D as the state, and the elements $Q_{{s_l}, {a_m}} (l, m =1, 2)$ in the Q-tables are randomly assigned a value between $(0,1)$, indicating that individuals are initially unfamiliar with the environment.  At round $t$, given the state $s$:
(i) With a probability $\epsilon$, each player randomly chooses an action $a\in\mathbb{A}$ to conduct trial-and-error exploration; otherwise, each chooses an action $a$ according to one's Q-table (i.e., $a$ is selected if $Q_{s,a}>Q_{s,\hat{a}}$).
(ii) Then, two players play the PD game and get a payoff $\pi$ according to the matrix Eq.~(\ref{eq:payoff}).
(iii) They get their new state $s'$ and update their Q-tables. Specifically,
the element $Q_{s,a}(t)$ just referred is updated as follows:
\begin{equation}
\begin{aligned}
Q_{s, a}(t+1) &=Q_{s, a}(t)+\alpha\left(\pi(t)+\gamma \max _{a^{\prime}} Q_{s^{\prime}, a^{\prime}}(t)-Q_{s, a}(t)\right) \\
&=(1-\alpha) Q_{s, a}(t)+\alpha\left(\pi(t)+\gamma \max _{a^{\prime}} Q_{s^{\prime}, a^{\prime}}(t)\right),
\end{aligned}
\end{equation}
where $\alpha\in(0,1]$ is the learning rate, which captures the contribution of the current step. A larger $\alpha$ means that the player is more forgetful, as old Q-values tend to be more rapidly modified.
$\pi(t)$ is the payoff obtained at present round following the payoff matrix Eq.~(\ref{eq:payoff}).
$\gamma\in[0,1)$ is the discount factor, measuring the weight of future rewards, as $\max _{a^{\prime}} Q_{s^{\prime}, a^{\prime}}(t)$ is the maximal value expected within the new state. The r.h.s. of the above equation indicates that the Q-values simultaneously contain the contribution of past experiences, reward at present and from the future.

The above process [steps (i)-(iii)] is repeated until the system reaches an equilibrium or the desired duration is completed. The three learning parameters are fixed at typical values of $\epsilon = 0.01$, $\alpha = 0.1$, $\gamma = 0.9$ throughout the study, where players appreciate both past experiences and expected rewards in the future.

\section{Results}\label{sec:results}
We report the evolution of cooperation for the three information schemes, where discontinuous transitions and bistability are uncovered, see Fig.~\ref{fig:phasetransition}. As shown in Fig.~\ref{fig:phasetransition}(a), when players focus on the opponent's action information (Scheme I), cooperation exhibits strong instability even at small values of temptation $b$. With the increase of $b$, the system evolves to a stable state dominated by mutual defection $f_c\approx0$. Correspondingly, the probability density function (PDF) curves of $f_c$ within the unstable interval in Fig.~\ref{fig:phasetransition}(d) show a trimodal distribution. With increasing $b$, the peaks at $0.5$ and $1$ both reduce.

By contrast, when players focus on their own action information (Scheme II), Fig.~\ref{fig:phasetransition}(b) shows that the mutual cooperation ($f_c\approx1$) is stable when $b\lesssim0.22$.
Further increasing $b$, however, leads to a dramatically different outcome --- the system either evolves into mutual cooperation for some experiments, or the system evolves into mutual defection for some other realizations, depending on the initial conditions. Once mutual cooperation or defection is reached, the later evolution of $f_c$ becomes quite stable, see the inset in Fig.~\ref{fig:phasetransition}(b).
When $b> b_c\approx0.32$, mutual defection is the only stable state.
The observation of bistable state is strengthened by the bimodal PDF as shown in Fig.~\ref{fig:phasetransition}(e). As expected, the peak of the mutual cooperation shrinks when $b$ is increased, while the peak of mutual defection goes up. These features indicate that there is a first-order-like phase transition for the cooperation prevalence in Scheme II.

Finally, when the two players are of asymmetric information structure (Scheme III), a similar phase transition and a bimodal PDF are observed, see Fig.~\ref{fig:phasetransition}(c,f). Yet, there is an essential difference compared to Scheme II that the cooperation prevalence $f_c$ shows a bounce between full cooperation and full defection, as shown in the inset of Fig.~\ref{fig:phasetransition}(c).
In addition, detailed examination shows that when the value of $b$ is larger, the possibility of cooperation emergence under Scheme III is higher than the value in Scheme II. For example, when $b=0.3$, $f_c \approx 0.25$ in Scheme III while $f_c \approx 0.15$ in Scheme II.

These results suggest that the information structure has a huge impact on the evolution of cooperation, and asymmetric information leads to new complexities in the form of first-order-like phase transition and true bistability.

\begin{figure}[tbp]
\centering
\includegraphics[width=1.0\linewidth]{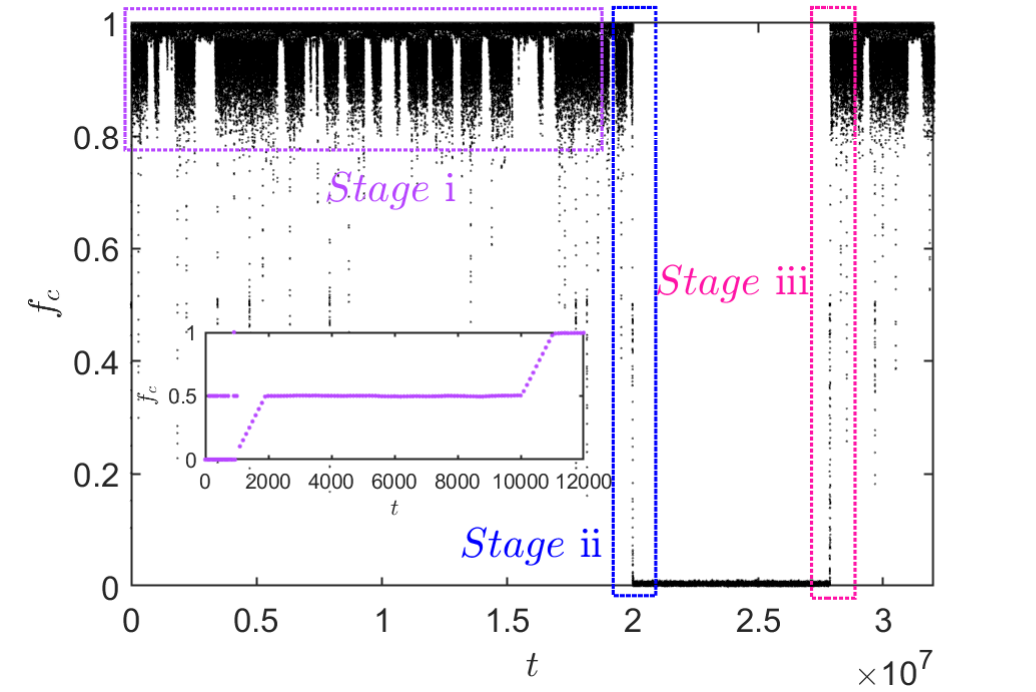}
\caption{\textbf{Typical time series of cooperation preference $f_c$ in Scheme \uppercase\expandafter{\romannumeral3}}. A sliding window average of 500 steps is conducted. Based on the characteristics displayed in the time series, it can be divided into three stages: i) Emergence of cooperation, ii) Breakdown of cooperation, and iii) Rebuilding of cooperation. The inset shows the time series of $f_c$ for the first $1.2\times10^4$ steps.
Parameters: $\epsilon$ = 0.01, $\alpha$ = 0.1, $\gamma$ = 0.9, $b$ = 0.2.
}
\label{fig:TimeSeries}
\end{figure}

\section{Mechanism Analysis}\label{sec:analysis}

\begin{figure}[tbph]
\centering
\includegraphics[width=1.0\linewidth]{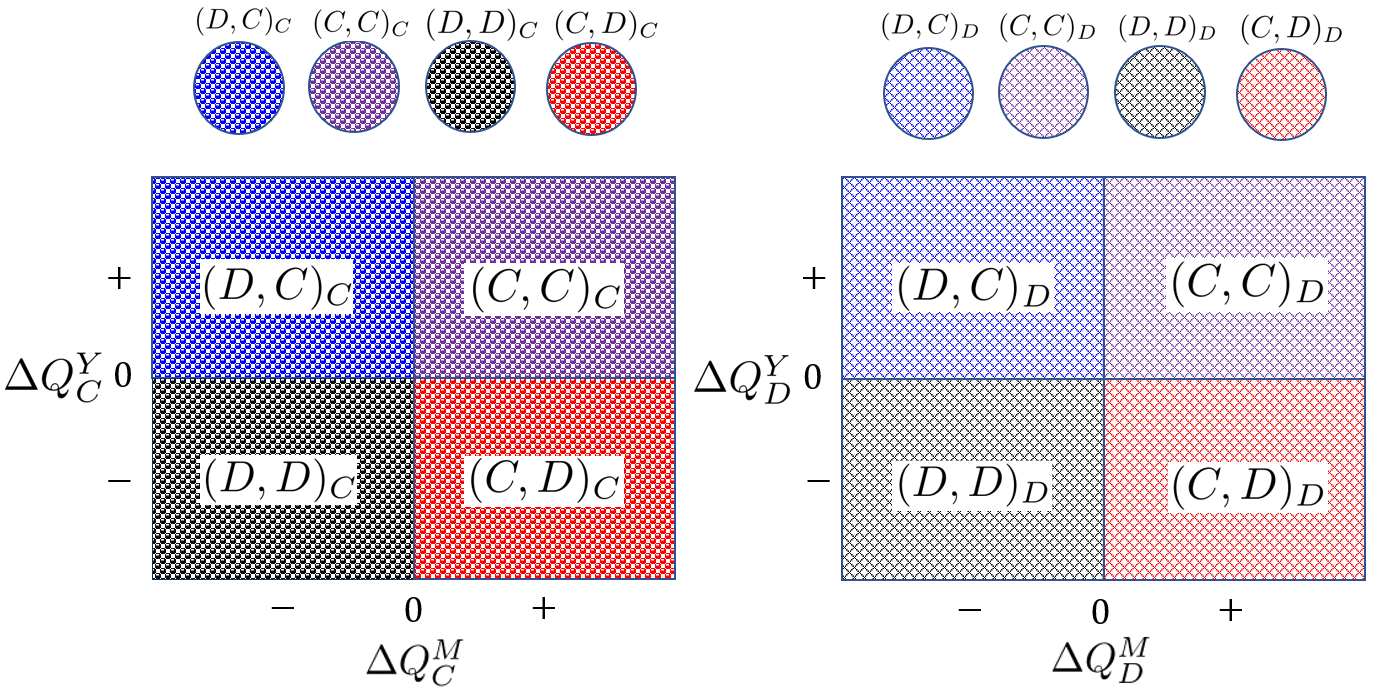}
\caption{\textbf{The action preference combinations of two players within two states.}
The four quadrants, based on the sign of the value $\Delta Q_{s_l}^{i}$ ($i\in\{M,Y\}$), represent the four possible combinations of action preferences in different states, denoted with subscripts. The left and right figures correspond to the system being in state C and state D, respectively. For example, the combination $(D, C)_D$ indicates that in state D, individual \emph{M} prefers action D, while individual \emph{Y} prefers action C.
}
\label{fig:preferencrCombination}
\end{figure}
Here, we primarily analyze the mechanisms under the asymmetric scenario in Scheme III. The mechanism analyses for Schemes I and II are relatively straightforward and are provided in Appendices~\ref{sec:appendixB} and ~\ref{sec:appendix}, respectively.

\begin{figure*}[t]
\centering
\includegraphics[width=1.0\linewidth]{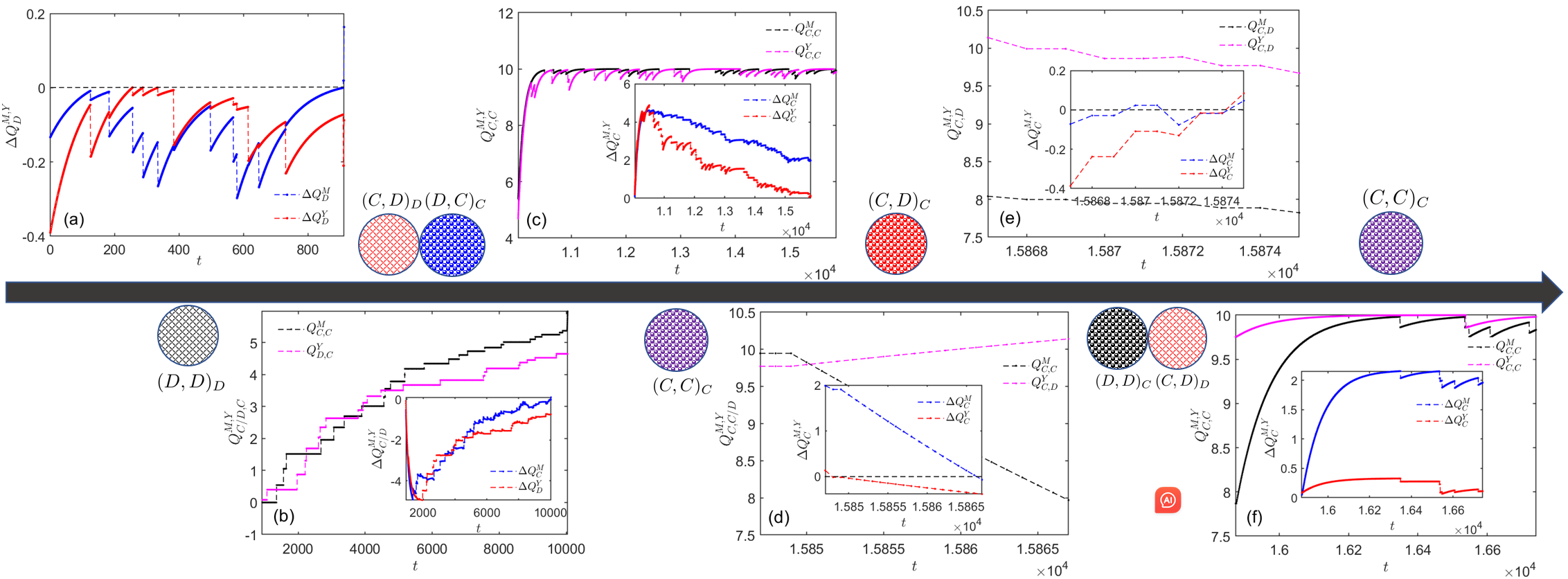}
\caption{\textbf{Cooperation emergence in Stage i.}
It shows the evolution of action combination preferences, and the temporal evolution of $Q_{s_l,a_m}^{M,Y}$ or $\Delta Q_{s_l}^{M,Y}$ values. Here, the action preference combination $(C,D)_D$$(D,C)_C$ indicates that individual \emph{M} chooses defect in state C and cooperate in state D, causing the system to cycle between $(C,D)_D$$\leftrightarrow$$(D,C)_C$. The same applies to $(D,D)_C$$(C,D)_D$. The sudden declines in (a) are because of occasional cooperation by exploration, where the action of defection brings to a reward $\pi=1+b$.
Parameters: $\epsilon$ = 0.01, $\alpha$ = 0.1, $\gamma$ = 0.9.
}
\label{fig:StageOne}
\end{figure*}
To understand the mechanism in the case of information asymmetry, we now turn to the evolution of the Q-table. To be certain, we categorize the evolutionary process into three stages based on the characteristics exhibited by the typical time series of $f_c$ shown in Fig.~\ref{fig:TimeSeries}, with questions as follows:
\begin{enumerate}
\item[1)] Stage i: how does cooperation emerge?
\item[2)] Stage ii: why does cooperation collapse?
\item[3)] Stage iii: how does cooperation reestablish afterwards?
\end{enumerate}

In addition to the elements $Q^i_{s_l,a_m}$ for each player $i$, we are particularly interested in their relative magnitude within a given row, i.e., $\Delta Q_{s_l}^{i}=Q_{s_l,a_1}^{i}-Q_{s_l,a_2}^{i}$. This value determines which action is preferred for player $i$ within the given state $s_l$. For example, if $\Delta Q_{s_l}^{i} \textgreater 0$, this means that for player $i$, the action C is preferred within the state $s_l$, otherwise D is supposed to be a better choice.
Accordingly, we explicitly show the action preference combinations within two states [see Fig.~\ref{fig:preferencrCombination}] based on the sign of $\Delta Q_{s_l}^{i}$, where $i\in\{M,Y\}$ labels the individual who considers their own action information (``Me'') and  the individual who considers the opponent's action information (``You''). For example, the action preference combination $(D,C)$ represents individual \emph{M} choosing action D and individual \emph{Y} choosing action C, which are denoted by different subscripts in different states: state C is represented by $(D, C)_C$, and state D by $(D, C)_D$.

To be certain, we start with a typical initial condition that is far from mutual cooperation $Q_{C,C}^{M}<Q_{C,D}^{M}$, $Q_{D,C}^{M}<Q_{D,D}^{M}$, $Q_{C,C}^{Y}>Q_{C,D}^{Y}$, $Q_{D,C}^{Y}<Q_{D,D}^{Y}$ and analyze the dynamical mechanisms. For other cases of total betrayal, refer to the evolution process in Stage iii.

\subsection {Stage i --- \emph{Cooperation emergence}}

To provide a clear and intuitive description of the evolutionary process at this stage, we divide the evolutionary mechanism of this stage into five distinct sub-stages.

\textbf{Sub-stage i} -- \emph{Two novices both prefer defection resulting in the preference combination of $(D,D)_D$.}

At the beginning, both players are unfamiliar with the environment, thus they prioritize immediate payoffs and learn that D is more beneficial, leading to $\Delta Q_{D}^{M,Y}<0$. Therefore, both exhibit self-interested behavior in the form of mutual defection $(D,D)_D$. However, the action preference combinations of $(D,D)_D$ bring very low payoffs to both parties, which weakens the advantage of choice D and cause both values of $\Delta Q_{D}^{M,Y}$ back to zero [Fig.~\ref{fig:StageOne}(a)]. The intermittent decreases are due to the action of C by exploration. When one of the values of $\Delta Q_{D}^{M,Y} \rightarrow 0$, their preferences in D are about to change.

\textbf{Sub-stage ii} -- \emph{Player M's action preference shift leads to a new action preference combination $(C, D)_D$ $\leftrightarrow$ $(D, C)_C$.}

When $\Delta Q_{D}^{M}> 0$ [Fig.~\ref{fig:StageOne}(a)], the player \emph{M}'s action preference is shifted from D to C. Correspondingly, the state of the system also undergoes the same change, and the system then enters a new action preference combination $(C, D)_D$ $\leftrightarrow$ $(D, C)_C$ [Fig.~\ref{fig:StageOne}(b)]. However, this action preference combination fails to persist in the presence of exploration.

\textbf{Sub-stage iii} -- \emph{Exploratory behavior of both parties favors cooperation and mutual cooperation $(C, C)_C$ is formed.}

Within the action preference combination $(C, D)_D$ $\leftrightarrow$ $(D, C)_C$, the exploratory behavior of both parties is conducive to the growth of the utility function $Q_{{s_l},{C}}$ [Fig.~\ref{fig:StageOne}(b)], and the values of $Q_{C,C}^{M}$ and $Q_{D,C}^{Y}$ increase discontinuously. Correspondingly, $\Delta Q_{C}^{M}$ and $\Delta Q_{D}^{Y}$ show an increasing trend [see inset in Fig.~\ref{fig:StageOne}(b)], indicating a gradual shift towards cooperation.
Due to asymmetric information causing a faster increase in $\Delta Q_{C}^{M}$ [see Appendix~\ref{sec:appendixA}], individual \emph{M} first transitions to cooperation in state D, leading the system to enter state C and establishing a stable positive feedback loop of mutual cooperation. As a result, the system enters a new action preference combination $(C, C)_C$, the value of $Q_{C,C}^{M,Y}$ remains unchanged after continuous rise in Fig.~\ref{fig:StageOne}(c).

\textbf{Sub-stage iv} -- \emph{Asymmetric information leads to exploitation of individual M by individual Y, the action preference combination $(C, D)_C$ is formed.}

The action preference combination $(C, C)_C$ remains unstable. The exploration behavior -- defection of both players leads to an increase in $Q_{C,D}^{M,Y}$. However, due to asymmetric information, $Q_{C,D}^{Y}$ increases more rapidly, and $\Delta Q_{C}^{Y}$ is falling at a faster rate than $\Delta Q_{C}^{M}$ [see inset in Fig.~\ref{fig:StageOne}(c)]. For more details, see Appendix~\ref{sec:appendixA}. Consequently, player \emph{Y} transitions from cooperation to defection in state C first, leading the system enter the action preference combination $(C, D)_C$. This combination can be viewed as a process of exploitation and tolerance. For individual \emph{M}, positive feedback from prior mutual cooperation results in $\Delta Q_{C}^{M}>0$, making \emph{M} inclined to cooperate even when faced with defection, showing tolerance. Thus, individual \emph{Y} can exploit \emph{M} by choosing defection for a period.

\textbf{Sub-stage v} -- \emph{Player M implements a punishment-like policy on player Y, the corresponding action preference combination is $(D,D)_C$ $\leftrightarrow$ $(C,D)_D$.}

However, tolerance within the action preference combination $(C, D)_C$ is limited. Frequent exploitation by the opponent causes a continuous decline in $Q_{C,C}^{M}$ [Fig.~\ref{fig:StageOne}(d)] and $\Delta Q_{C}^{M}$ to show a decreasing trend [see inset in Fig.~\ref{fig:StageOne}(d)]. When $\Delta Q_{C}^{M} < 0$, individual \emph{M} switches from action C to D in state C, transitioning the system to the combination preference of $(D,D)_C$ $\leftrightarrow$ $(C,D)_D$. Within this combination, individual \emph{Y}'s persistent exploitation from the previous sub-stage becomes intermittent, resulting in a reduced payoff and causing $Q_{C,D}^{Y}$ to start declining [Fig.~\ref{fig:StageOne}(e)]. This can be seen as a punishment process by individual \emph{M} towards individual \emph{Y}. This process causes $\Delta Q_{C}^{Y}$ to rise [see inset in Fig.~\ref{fig:StageOne}(e)]. When $\Delta Q_{C}^{Y} > 0$, individual \emph{Y} reverts to cooperation, forming a positive feedback loop that returns the system to $(C,C)_C$.

\begin{figure}[bpth]
\centering
\includegraphics[width=0.9\linewidth]{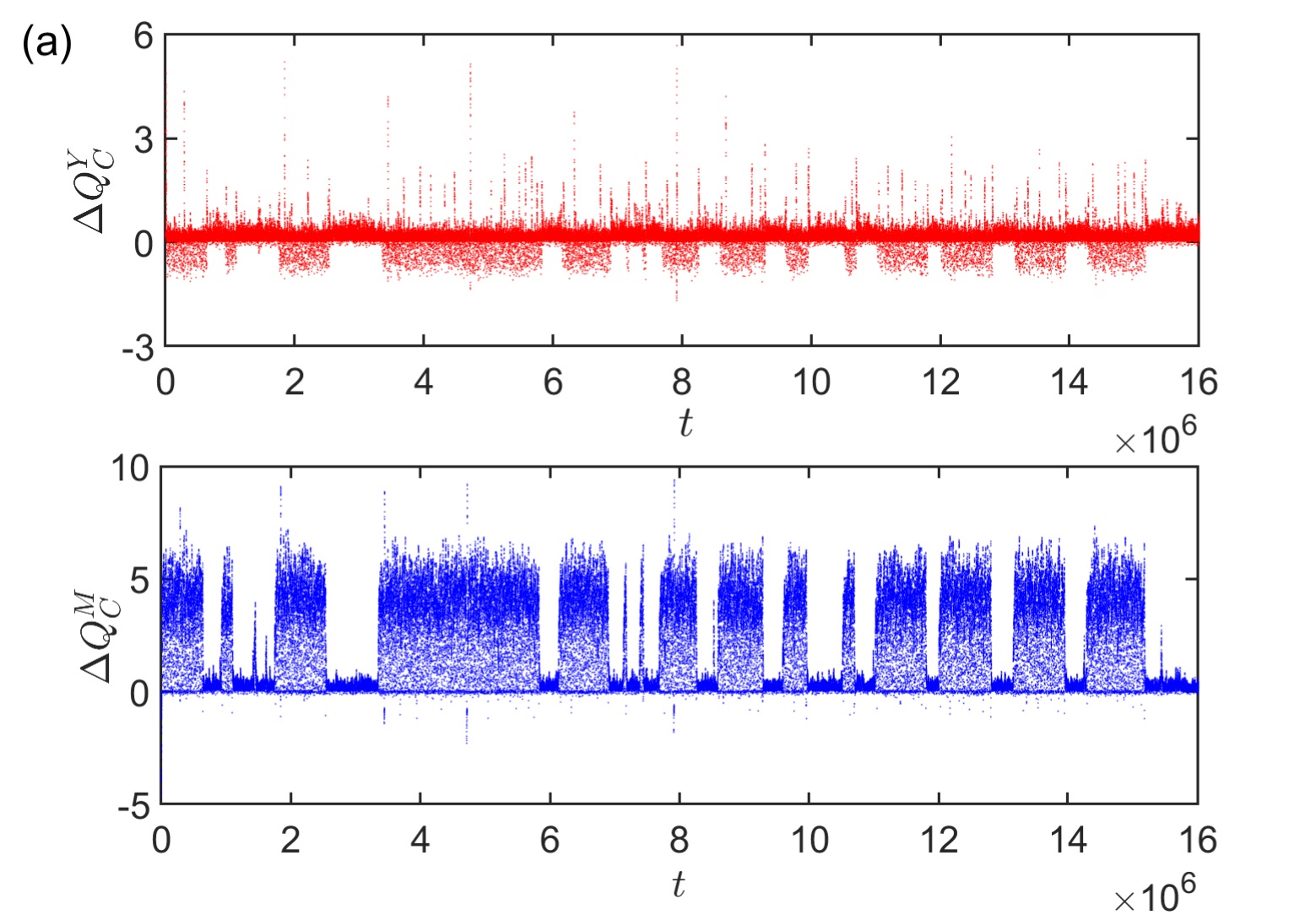} \\ 
\includegraphics[width=0.9\linewidth]{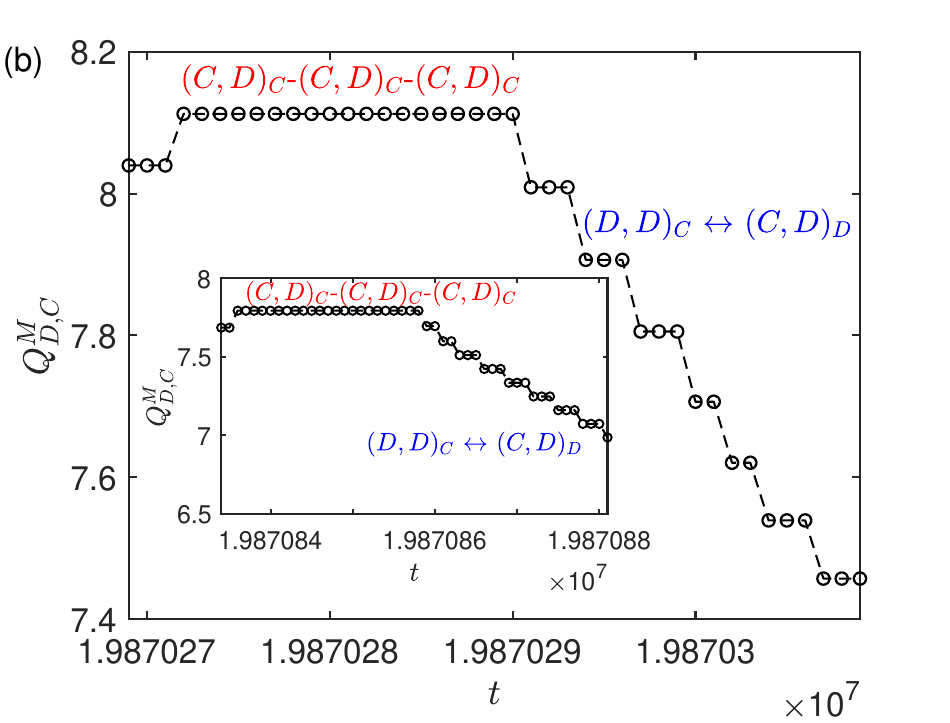}
\caption{\textbf{Cooperation collapse in Stage ii}.
(a) The time evolution of $\Delta Q_{C}^{Y,M}$. The upper panel shows the time evolution of $\Delta Q_{C}^{Y}$, where intermittent occurrences of $\Delta Q_{C}^{Y}<0$ can be observed, indicating the accumulation of exploitation of individual \emph{M} by individual \emph{Y}. The lower panel shows the evolution of $\Delta Q_{C}^{M}$, with corresponding intermittent oscillations observed in the upper panel, each oscillation representing a punishment process of individual \emph{Y} by individual \emph{M}.
(b) The time evolution of $Q_{D,C}^{M}$. It can be observed that the decrease in $Q_{D,C}^{M}$ mainly occurs during the punishment process of individual \emph{Y} by individual \emph{M}, with the corresponding action preference combination being $(D,D)_C$$\leftrightarrow$$(C,D)_D$. The inset shows the evolution of $Q_{D,C}^{M}$ over different time periods.
Parameters: $\epsilon$ = 0.01, $\alpha$ = 0.1, $\gamma$ = 0.9.
}
\label{fig:StageTwo}
\end{figure}

Stage i shows the evolution of cooperation emergence -- exploitation and tolerance -- punishment -- mutual cooperation. However, the completion of this stage does not establish a stable cooperative relationship between the two players. As shown in the top panel of Fig.~\ref{fig:StageTwo}(a), the condition $\Delta Q_C^{Y} < 0$ occurs intermittently, indicating that individual \emph{Y} still exploits individual \emph{M} from time to time. As a result, the process of sub-stages iii-v intermittently occurs in the subsequent evolution, causing $\Delta Q_C^{M}$ to fluctuate in the bottom panel of Fig.~\ref{fig:StageTwo}(a). Despite this, the system maintains a relatively high average cooperation preference ($f_c \textgreater 0.8$) until it eventually transitions to a state of complete defection. This outcome is attributed to individual \emph{M}'s inclination to cooperate in state D.

\subsection {Stage ii --- \emph{Cooperation collapse}}

In Stage i, we observe that individual \emph{M} frequently ``forgives'' individual \emph{Y} and reestablished mutual cooperation. However, an intriguing phenomenon emerges afterwards: individual \emph{M} gradually loses patience and is no longer inclined to cooperate. As shown in Fig.~\ref{fig:StageTwo}(b) and the inset, the decline in $Q_{D,C}^{M}$ primarily occurs during sub-stage V. This indicates that each time individual \emph{M} punishes individual \emph{Y}, \emph{M}'s inclination to choose cooperation in state D diminishes. Once tolerance is completely eroded, the system transitions into state D, resulting in a collapse of cooperation. Consequently, when the opponent exploits again, the system shifts to a state of mutual defection $(D,D)_D$.

\begin{figure*}[bpth]
\centering
\includegraphics[width=1.0\linewidth]{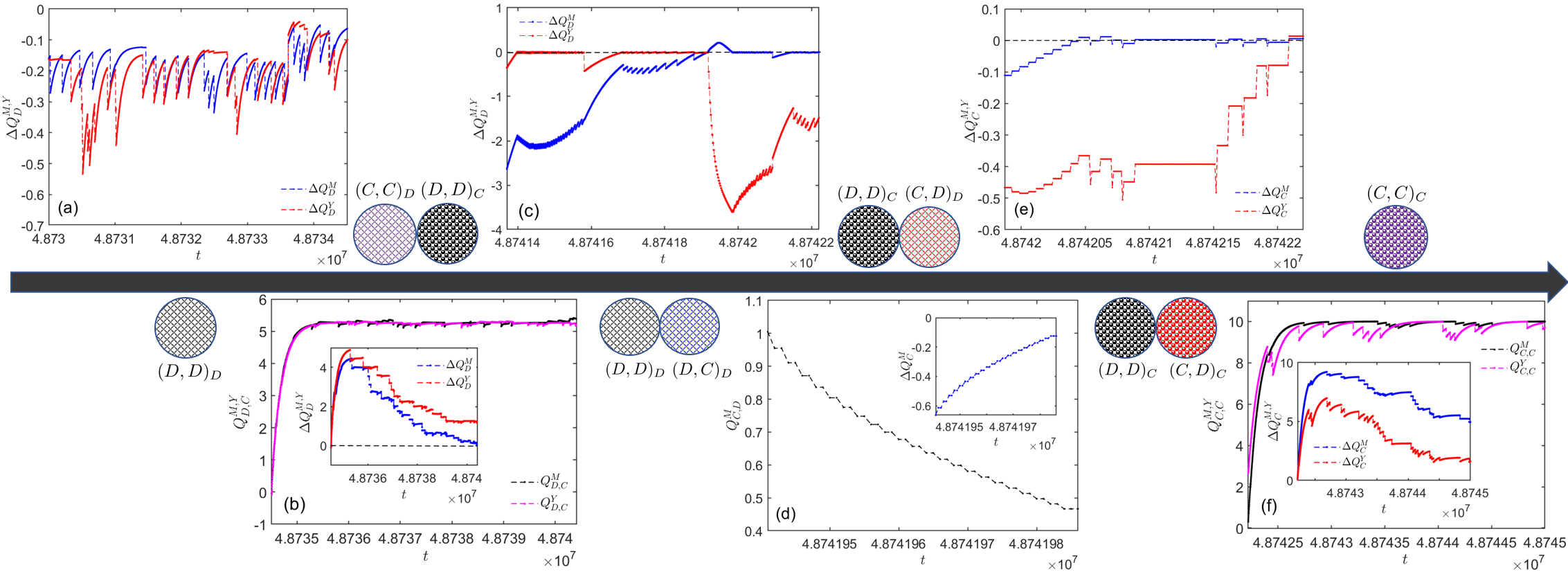}
\caption{\textbf{Cooperation reestablishment in Stage iii}. It shows the evolution of action combination preferences, and the temporal evolution of $Q_{s_l,a_m}^{M,Y}$ or $\Delta Q_{s_l}^{M,Y}$. Here, the action preference combination $(C,C)_D$ $(D,D)_C$ indicates that individual \emph{M} chooses defect in state C and cooperate in state D, causing the system to cycle between $(C,C)_D$$\leftrightarrow$$(D,D)_C$. The same applies to $(D,D)_C$$(C,D)_D$. The action pair $(D,D)_D$$(D,C)_D$ indicates that in state D, individual \emph{Y} alternates between cooperation and defection. Similarly, $(D,D)_C (C,D)_C$ shows that individual \emph{M} switches between cooperation and defection in state C.
The sudden declines in (a) are because of occasional cooperation by exploration, where the action of defection brings to a reward $\pi=1+b$.
Parameters: $\epsilon$ = 0.01, $\alpha$ = 0.1, $\gamma$ = 0.9.
}
\label{fig:StageThree}
\end{figure*}

\begin{figure}[bpth]
\centering
\includegraphics[width=1.0\linewidth]{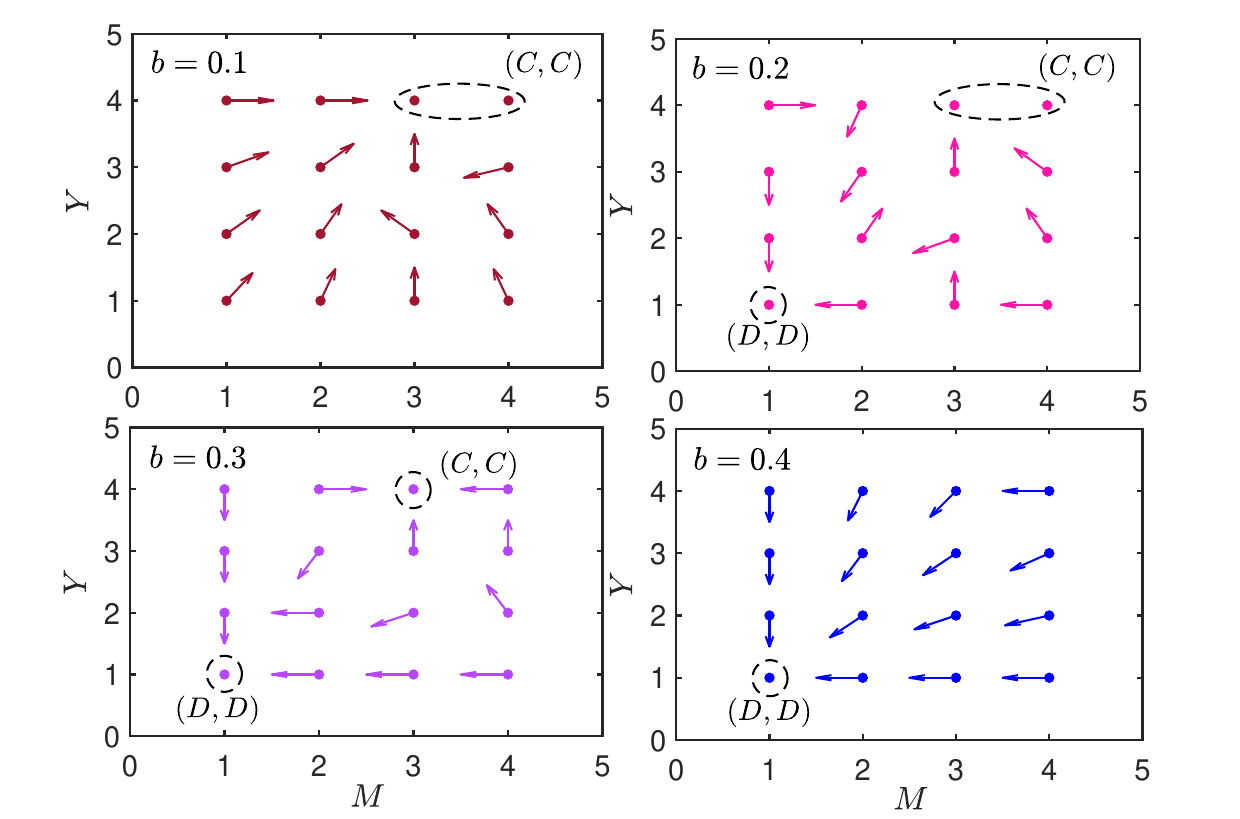}
\caption{\textbf{Evolutionary paths in Scheme III.} Starting with all possible settings of the initial Q-table for the two players (labeled ``\emph{Y}'' and ``\emph{M}'') for four typical dilemma intensities $b$. The axis labels $1-4$ respectively represent combinations of ($\Delta Q_{C}<0$, $\Delta Q_{D}<0$), ($\Delta Q_{C}<0$, $\Delta Q_{D}>0$), ($\Delta Q_{C}>0$, $\Delta Q_{D}<0$), ($\Delta Q_{C}>0$, $\Delta Q_{D}>0$). The arrows show the evolutionary directions of the combination type during a fixed time interval $t=3\times10^5$.
Parameters: $\epsilon$ = 0.01, $\alpha$ = 0.1, $\gamma$ = 0.9.
}
\label{fig:Attractor}
\end{figure}

\subsection {Stage iii --- \emph{Cooperation reestablishment}}

There the system transitions away from $(D,D)_D$ to $(C,C)_C$ again. Three distinct sub-stages can be divided in this stage.

\textbf{Sub-stage i} -- \emph{Simultaneous cooperative exploration breaks mutual defection, triggers $(C,C)_D$ $\leftrightarrow$ $(D,D)_C$ cyclic state.}

Within the mutual defection state, the payoff $\pi$ for either is zero, reducing their preference in defection. This is evidenced by the upward trend in $\Delta Q_{D}^{M,Y}$ shown in Fig.~\ref{fig:StageThree}(a), the intermittent declines are due to occasional cooperative actions during exploration. Unilateral cooperation, however, only strengthens the other player's preference for defection because their preference in state C remains defection (i.e., $\Delta Q_{C}^{M,Y} < 0$). Simultaneous cooperation by both players can alter this situation. When both choose to cooperate, they each receive a payoff $\pi = R$, which triggers an increase in $Q_{D,C}^{M,Y}$ and leads to $\Delta Q_{D}^{M,Y} > 0$, indicating a reversal in preference as shown in Fig.~\ref{fig:StageThree}(b). The system then enters a cyclical state of $(C,C)_D$ $\leftrightarrow$ $(D,D)_C$. However, this action preference combination cannot be sustained under weak exploration.

\textbf{Sub-stage ii} -- \emph{Alternating exploitation and punishment prepare for reestablishing cooperation.}

Within the action preference combination $(C,C)_D$ $\leftrightarrow$ $(D,D)_C$, the exploration behavior -- defection of both players leads to an increase in $Q_{D,D}^{M,Y}$. Due to asymmetric information, $Q_{D,D}^{M}$ increases more rapidly (for more details, see Appendix ~\ref{sec:appendixA}), causing $\Delta Q_{D}^{M}$ to decrease faster than $\Delta Q_{D}^{Y}$ [see inset in Fig.~\ref{fig:StageThree}(b)].
Consequently, player \emph{M} transitions from cooperation to defection in state D first, leading the system enter the action preference combination $(D, C)_D$ -- a process is similar to the exploitation and tolerance observed in sub-stage iv of stage i, with the roles reversed: \emph{M} exploits \emph{Y}, while \emph{Y} tolerates \emph{M}.

Then, player \emph{Y} implements a similar punishment-like strategy on player \emph{M}. Within the action preference combination $(D, C)_D$, \emph{M}'s continuous exploitation leads to a persistent decline in $Q_{D,C}^{Y}$. When $\Delta Q_{D}^{Y}\rightarrow 0$, \emph{Y} gains no advantage in choosing either cooperation or defection, causing $\Delta Q_{D}^{Y}$ to fluctuate around zero [Fig.~\ref{fig:StageThree}(c)]. The corresponding action preference combination is $(D,D)_D$ $\leftrightarrow$ $(D,C)_D$, which then predominantly shifts to mutual defection $(D,D)_D$ $\leftrightarrow$ $(D,D)_D$. This process results in an increase in $\Delta Q_{D}^{M}$.

When $\Delta Q_{D}^{M}>0$, individual \emph{M} re-chooses cooperation in state D.  The roles of \emph{M} and \emph{Y} then reverse,  repeating the previously described process. During this period, fluctuations of $\Delta Q_{D}^{M}$ around zero occasionally revert the system to state C, leading to a decrease in $Q_{C,D}^{M}$ [Fig.~\ref{fig:StageThree}(d)]. When $\Delta Q_{C}^{M}> 0$, the player \emph{M}'s action preference in state C shifts from defection to cooperation.  However, due to the lack of positive returns, $\Delta Q_{C}^{M}$ fluctuates around zero [Fig.~\ref{fig:StageThree}(e)]. This punishment-like strategy on player \emph{Y} again results in an increase in $\Delta Q_{C}^{Y}$. When $\Delta Q_C^{Y}>0$, \emph{Y}'s action preference in state C also shifts towards cooperation. This indicates that once the system returns to state C, the positive feedback from mutual cooperation can reestablish cooperation between both parties.

\begin{figure*}[t]
\centering
\includegraphics[width=0.33\linewidth]{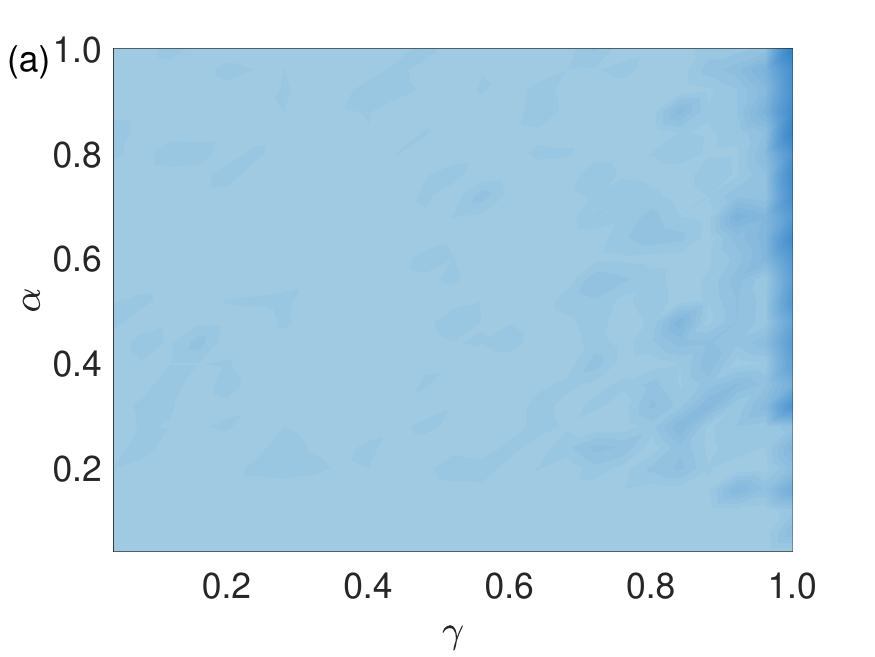}
\includegraphics[width=0.33\linewidth]{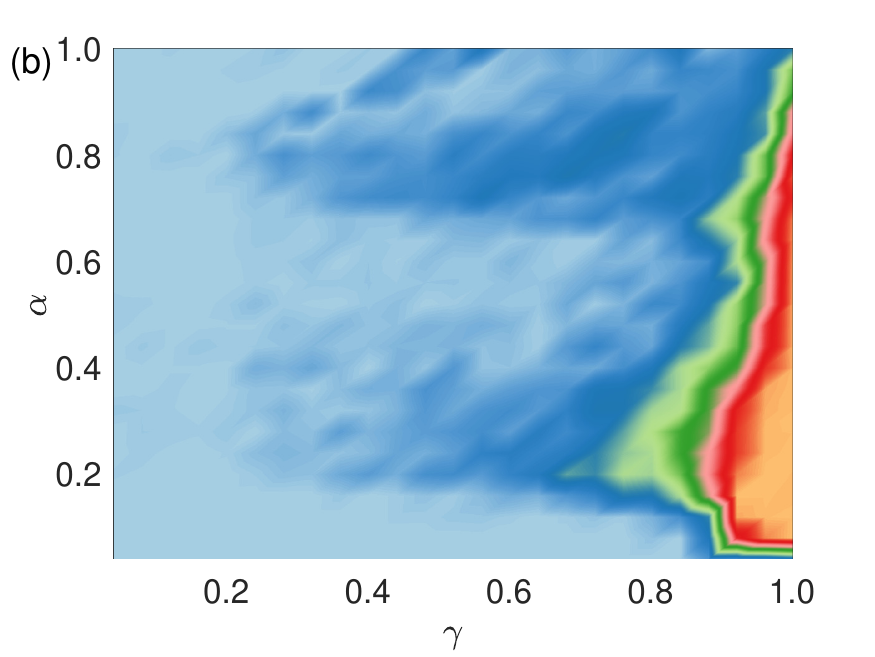}
\includegraphics[width=0.33\linewidth]{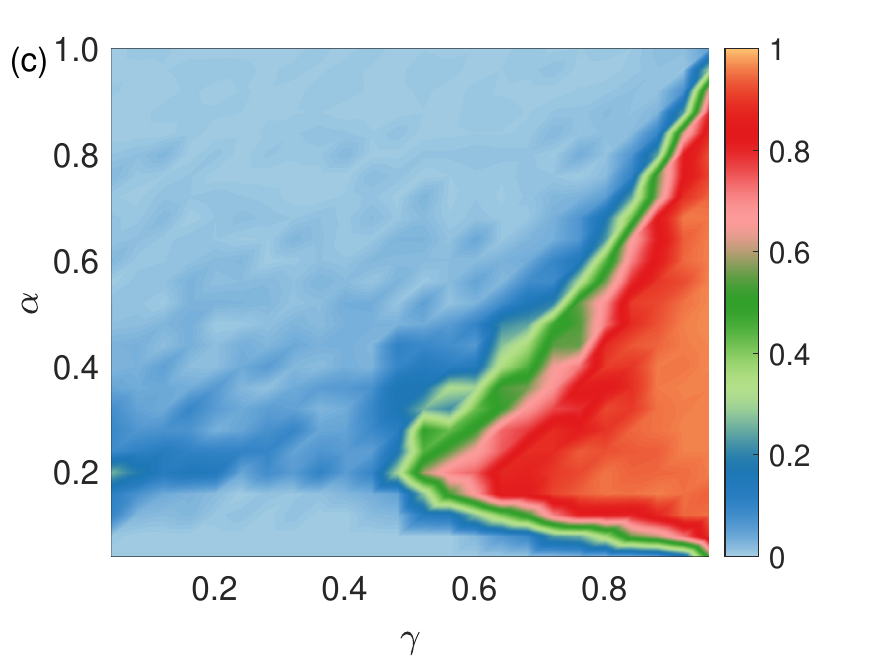}
\caption{(Color online) \textbf{The color-coded averaged cooperation preference $f_c$ in the domain ($\gamma$, $\alpha$).}
 (a-c) are respectively for Scheme I-III. The red regions indicate that cooperation dominates, which often emerge for the combination of a small learning rate $\alpha$ and a large discount factor $\gamma$.
Each data is averaged 100 realizations, and for each realization the data is averaged 500 rounds after a transient of $2\times10^8$ steps.
Other parameters : $\epsilon$ = 0.01, $b$ = 0.2.
}
\label{fig:PhaseDiagram}
\end{figure*}

\begin{figure}[bpth]
\centering
\includegraphics[width=0.8\linewidth]{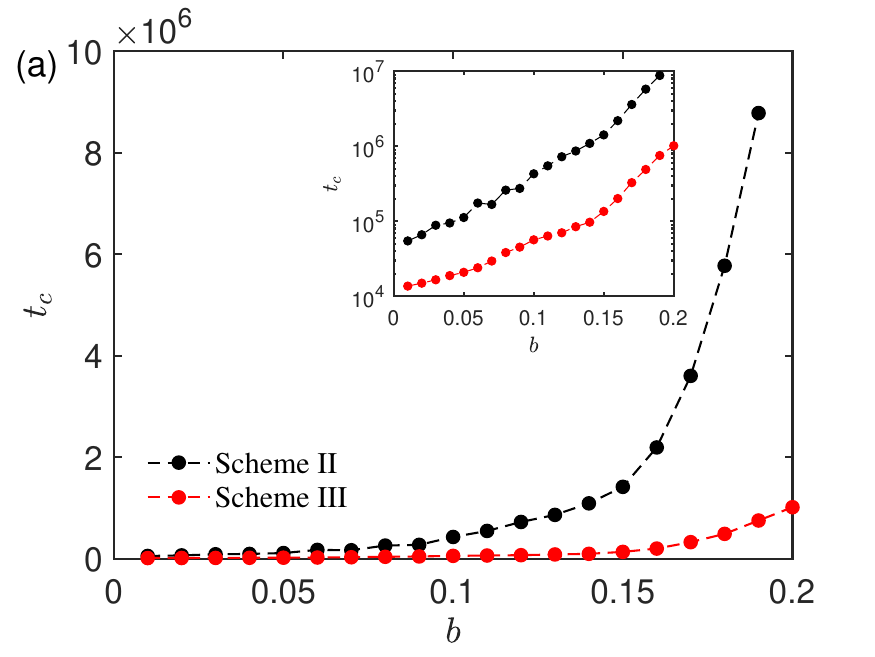}\\
\includegraphics[width=0.8\linewidth]{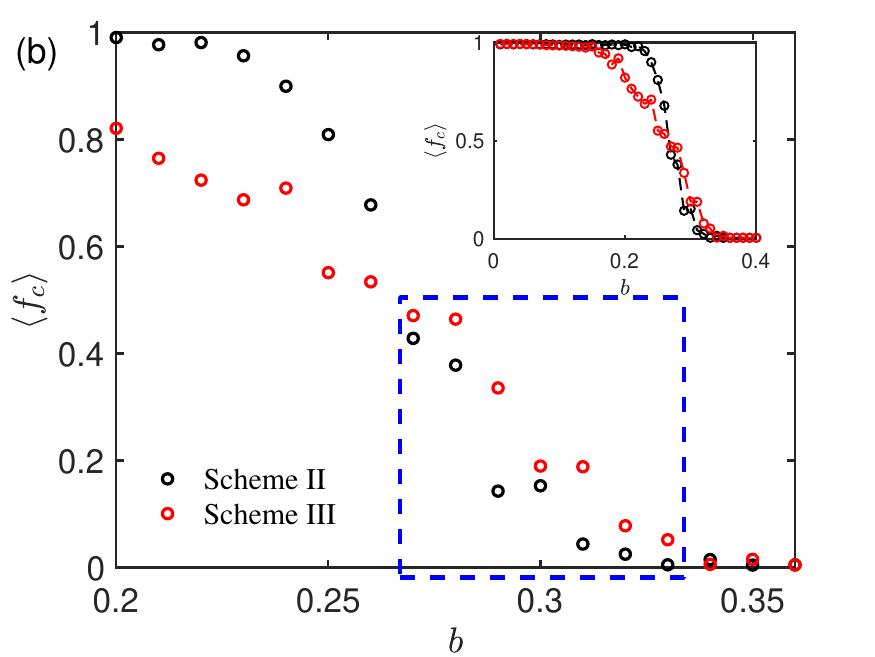}
\caption{\textbf{Comparison between Scheme II and III.}
(a) The convergence time $t_c$ versus the dilemma strength $b$, and the inset shows the same data but $y$-axis is taken logarithmic.
(b) The averaged cooperation preference $\langle f_c \rangle$ versus $b$, 100 ensemble averages are conducted for each data besides the time average as we did in Fig.~\ref{fig:phasetransition}(a-c).
Other parameters: $\epsilon$ = 0.01, $\alpha$ = 0.1, $\gamma$ = 0.9.
}
\label{fig:tcTowardsCooperation}
\end{figure}

\textbf{Sub-stage iii} -- \emph{Cooperation is reestablished when individual M chooses cooperation.}

Up to this point, the two individuals have reached a consensus to cooperate in state C. When individual \emph{M} re-chooses cooperation, the system enters state C and mutual cooperation is successfully reestablished.  In Fig.~\ref{fig:StageThree}(f), a result seemingly identical to that in Fig.~\ref{fig:StageOne}(c) indicates that the system returns to the Stage i evolution process.

Finally, to gain an intuitive understanding of the cooperation evolution in Scheme III, we show evolutionary paths for four typical dilemma strengths $b$, see Fig.~\ref{fig:Attractor}. These are obtained by the following procedures. Starting with all possible combinations of the two Q-tables (i.e., $4\times4$ cases), we monitor the evolution of these combinations, where some ``attractors'' are observed. For a small value of dilemma strength ($b=0.1$), mutual cooperation is the only stable solution, while for a large value ($b=0.4$) mutual defection is exclusively stable. For the cases in between ($b=0.2, 0.3$), the two attractors compete with each other, the evolution of the system is up to which basin of attraction are their initial conditions located.
The observations align with the overall picture discussed above.

\section{Further comparison}\label{sec:compare}

In this section, we first present the average cooperation preference $f_c$ in the domain of two key learning parameters ($\gamma$, $\alpha$) for three different schemes, with the dilemma strength fixed at $b=0.2$, as shown in Fig.~\ref{fig:PhaseDiagram}.
We find that in Scheme I, there is no emergence of cooperation across the region for the given $b$ [Fig.~\ref{fig:PhaseDiagram}(a)], instead decent levels of cooperation are observed in the other two schemes. In Fig.~\ref{fig:PhaseDiagram}(b, c), the red regions indicate that cooperation dominates ($f_c\sim 0.8$), where the learning rate $\alpha$ is mostly small and the discount factor $\gamma$ is large.
The observation can be interpreted as that a high level of cooperation emerges only when players both pay attention to their historical experiences and have a long-term vision. Besides, the region dominated by cooperation within Scheme III is wider than that of Scheme II. Detailed examination shows that, in the case of asymmetric information, a moderate degree of future expectation $\gamma$ is sufficient to trigger the emergence of cooperation given a small value of the learning rate $\alpha$.

Apart from the average cooperation preference $\langle f_c \rangle$ at the final state, the convergence time towards the final state also matters. Fig.~\ref{fig:tcTowardsCooperation}(a) shows that average convergence time $t_c$ for the system towards full cooperation are much shorter in Scheme III than the values within Scheme II. Across the whole range of $b$, the converge time in Scheme II is about one order larger compare to the case of Scheme III.
In Fig.~\ref{fig:tcTowardsCooperation}(b), we can observe that there is a crossover in the average cooperation preference as $b$ is varied.
A higher $\langle f_c \rangle$ in Scheme II is observed when $b< 0.26$, while the opposite observation is made when $b> 0.26$. The reason behind the difference shown in Fig.~\ref{fig:tcTowardsCooperation} is closed related to the evolutionary mechanism of Scheme II, which is analyzed in the Appendix~\ref{sec:appendix}.

\section{Discussion}\label{sec:discussion}


In summary, we explore the evolution of cooperation in the iterated prisoner's dilemma game under three distinct information scenarios within the reinforcement learning (RL) framework. Unlike existing studies, we focus on how different information perceptions influence cooperation dynamics. Our findings demonstrate that information structure plays a critical role: in symmetric scenarios, direct action-state associations foster cooperation, while the asymmetric scenario promotes faster and more robust cooperation emergence. The evolutionary dynamics exhibit first-order-like phase transitions, with cooperation preference oscillating between mutual cooperation and defection. Mechanism analysis reveals the processes of cooperation emergence, breakdown, and reconstruction, alongside identifying basins of attraction for stable states at specific dilemma intensities.


While most research focuses on the emergence and maintenance of cooperation~\cite{Roca2009Evolutionary,Sigmund2010The}, few address its breakdown and reconstruction~\cite{Ding2023Emergence}. Our study highlights that moderate tolerance can sustain cooperation, but excessive exploitation risks its collapse, aligning with real-world observations. Rebuilding cooperation is challenging, often leaving exploiters at a disadvantage.


This work is an initial step in understanding information perception's role in cooperation within RL. We limit our analysis to three simple information structures in two-player scenarios, but real-world complexities—such as diverse personal and societal factors~\cite{Henrich2004Foundations} and intricate social networks~\cite{Newman2018networks}—warrant further investigation. Additionally, integrating moral preference hypotheses~\cite{Capraro2021Mathematical} with RL to better simulate decision-making presents a promising future direction.

\begin{acknowledgments}
This work was supported by the National Natural Science Foundation of China [Grants Nos. 12075144,12165014] and Fundamental Research Funds for the Central Universities (GK202401002).
\end{acknowledgments}

\appendix

\section{Asymmetric information causes imbalanced Q-value evolution between two players}\label{sec:appendixA}

\subsection {Within sub-stage iii of stage i}
There the system in a  $(C,D)_D$ $\leftrightarrow$ $(D,C)_C$ cyclic state. Individual \emph{M} cooperates in state D and defects in state C, while individual \emph{Y} cooperates in state C and defects in state D.

For individual \emph{M}: exploratory cooperation in state C shifts the action preference from $(C,D)_D$ $\leftrightarrow$ $(D,C)_C$ to $(C,D)_D$ $\leftrightarrow$ $(C,C)_C$, resulting in an immediate mutual cooperation payoff of $R=1$ and an increase in $Q_{C,C}^M$. This exploratory behavior also drives the system to state C, yielding a temptation value of $T=1.2$ under the $(D,C)_C$ preference, thus accelerating the increase in $Q_{C,C}^{M}$.

For individual \emph{Y}: exploratory cooperation in state D shifts the action preference from $(C,D)_D$ $\leftrightarrow$ $(D,C)_C$ to $(C,C)_D$ $\leftrightarrow$ $(D,C)_C$, resulting in an immediate mutual cooperation payoff of $R=1$ and an increase in $Q_{D,C}^{Y}$. However, since individual \emph{Y} cannot directly alter the system's state, it continues along its previous trajectory into state C, where under the $(D,C)_C$ action preference, it receives the payoff for sucker, $S=-0.2$, without the additional incentive seen in individual \emph{M}.

\subsection {Within sub-stage iv of stage i}
There the system in a $(C,C)_C$ state, both individuals choose to cooperate in state C.

For individual \emph{M}: exploratory defection in state C drives the system to state D, shifting the action preference combination from $(C,C)_C$ to $(D,C)_C$ $\leftrightarrow$ $(C,D)_D$, then back to $(C,C)_C$ [consistent with the process in sub-stage iii]. Then, an immediate temptation payoff of $T=1.2$ is obtained and $Q_{C,D}^{M}$ is increased.

For individual \emph{Y}: exploratory defection in state C shifts the action preference combination from  $(C,C)_C$ to $(C,D)_C$, then back to $(C,C)_C$. This results in an immediate temptation payoff of $T=1.2$, leading to an increase in $Q_{C,D}^{Y}$. However, unlike individual \emph{M}, \emph{Y} cannot alter the system's state, thus bypassing the process of reverting to sub-stage iii, consequently accelerating the increase in $Q_{C,D}^{Y}$.

\subsection {Within sub-stage ii of stage iii}
There the system in a $(C,C)_D$ $\leftrightarrow$ $(D,D)_C$ cyclic state. Both individuals choose to cooperate in state D and defect in state C.

For individual \emph{M}: exploratory defection in state D shifts the action preference from $(C,C)_D$ $\leftrightarrow$ $(D,D)_C$ to $(D,C)_D$ $\leftrightarrow$ $(C,C)_D$, resulting in an immediate temptation payoff of $T=1.2$ and an increase in $Q_{D,D}^M$. This exploratory behavior also drives the system to state D, yielding a mutual cooperation payoff of $R=1$ under the $(C,C)_D$ preference, thus accelerating the increase in $Q_{D,D}^M$.

For individual \emph{Y}: exploratory defection in state D shifts the action preference from $(C,C)_D$ $\leftrightarrow$ $(D,D)_C$ to $(C,D)_D$ $\leftrightarrow$ $(D,D)_C$, resulting in an immediate temptation payoff of $T=1.2$ and an increase in $Q_{D,D}^Y$. However, since individual \emph{Y} cannot alter the system's state, it continues along its previous trajectory into state C, where under the $(D,D)_C$ action preference, it receives the payoff for punishment, $P=0$, without the additional incentive seen in individual \emph{M}.

\section{Mechanism analysis in Scheme I}\label{sec:appendixB}

\begin{figure}[bpth]
\centering
\includegraphics[width=0.9\linewidth]{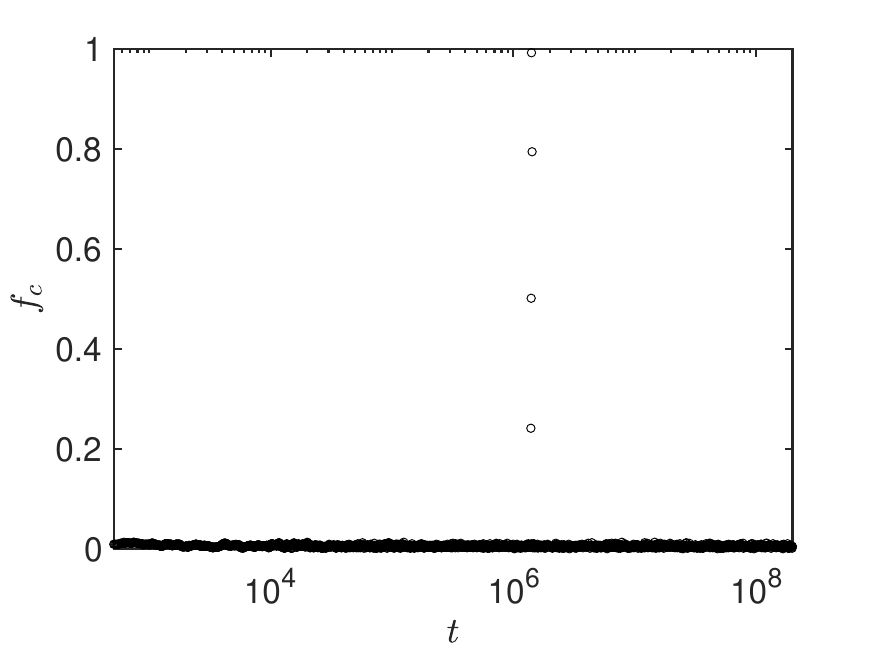}
\caption{Typical time series of cooperation preference $f_c$ in Scheme I. A sliding window average of 500 steps is conducted. As can be seen from the figure, cooperation fails to emerge and be sustained.
Parameters: $\epsilon$ = 0.01, $\alpha$ = 0.1, $\gamma$ = 0.9, $b$ = 0.2.}
\label{fig:TimeSeries_one}
\end{figure}

In Scheme I, both individuals focus on the opponent’s actions, resembling the Tit for Tat (TFT) strategy but falling short of fully implementing it. A key limitation is the difficulty in establishing and maintaining a ``cooperation-cooperation'' pattern. Moreover, since individuals cannot directly determine the state, the system lacks the ability to enforce punishment-like strategies, as seen in Schemes II and III. Even when starting from TFT-like initial conditions, cooperation proves unsustainable. Occasional misunderstandings gradually erode the tendency to cooperate in state C, ultimately leading to mutual defection.

The underlying evolutionary mechanism involves a key issue in maintaining mutual cooperation: both parties must always choose to cooperate in state C. Once one party shifts from cooperation to betrayal, the system will enter a stage of mutual exploitation. When exploratory betrayal behaviors accumulate advantages over time, the cooperation conditions will no longer be met, ultimately leading the system into a state of mutual betrayal. As can be seen from the Fig.~\ref{fig:TimeSeries_one}, cooperation fails to emerge and be sustained.

For more details, we denote two individuals as $i=\{Y_1, Y_2\}$, who consider their opponent's action information. The values of $Q_{s_l, a_m}$ and $\Delta Q_{s_l}^{i}$ are labeled as in the text. Even starting from an initial condition of full cooperation, i.e., $Q_{C,C}^{Y_1,Y_2} > Q_{C,D}^{Y_1,Y_2}$ and $Q_{D,C}^{Y_1,Y_2} > Q_{D,D}^{Y_1,Y_2}$, occasional exploratory choices of betrayal by both parties will lead to an increase in $Q_{C,D}^{Y_1,Y_2}$.

After the advantage of betrayal accumulates over time, satisfying $Q_{C,D}^{Y_1/Y_2} > Q_{C,C}^{Y_1/Y_2}$, one individual switches from cooperation to betrayal in state C. This initiates a continuous exploitation process $(C_D,D_C)$$-$$(C_D,D_C)$$-$$(C_D,D_C)$. As continuous exploitation causes the other individual's tendency to cooperate in state D to decline, they eventually switch to betrayal in state D, leading to another continuous exploitation process with roles reversed $(D_C,C_D)$$-$$(D_C,C_D)$$-$$(D_C,C_D)$. This process ultimately results in both individuals having no inclination to choose cooperation in either state, leading to the tragedy of total betrayal.

\section{Mechanism analysis in Scheme II}\label{sec:appendix}

\begin{figure}[bpth]
\centering
\includegraphics[width=0.88\linewidth]{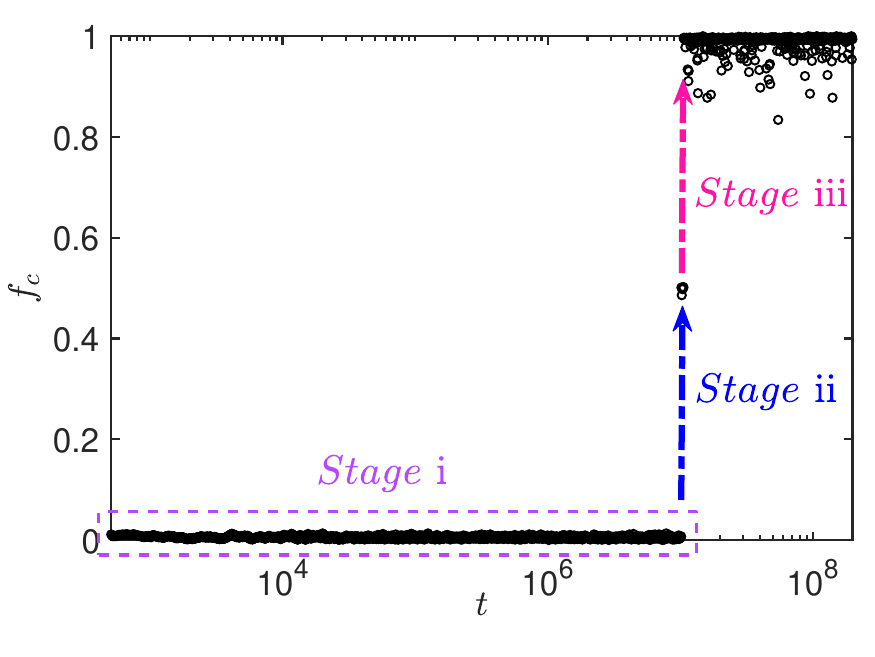}
\caption{Typical time series of cooperation preference $f_c$ in Scheme II. A sliding window average of 500 steps is conducted. Based on the characteristics displayed in the time series, it can be divided into three stages: i) Mutual betrayal, ii) Breaking away from mutual betrayal, and iii) Establishing of cooperation.
Parameters: $\epsilon$ = 0.01, $\alpha$ = 0.1, $\gamma$ = 0.9, $b$ = 0.2.
}
\label{fig:TimeSeries_two}
\end{figure}

\begin{figure*}[bpth]
\centering
\includegraphics[width=1.0\linewidth]{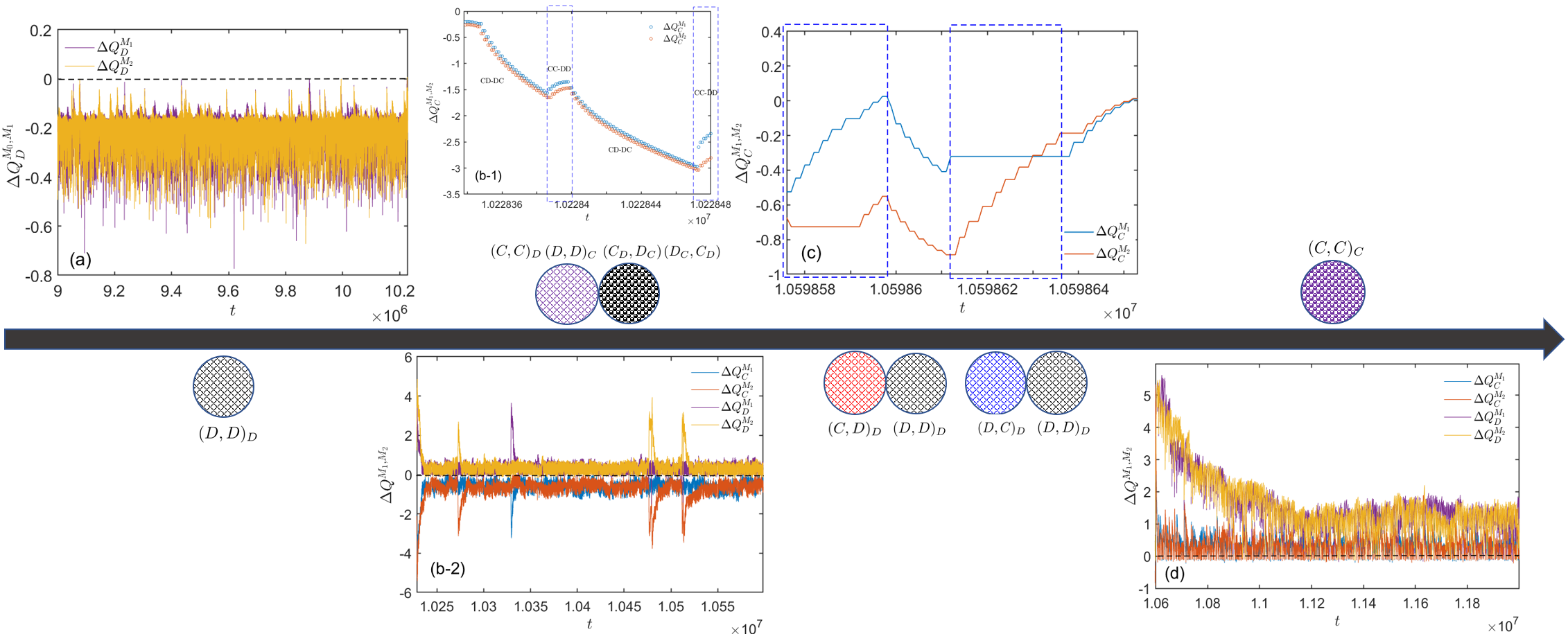}
\caption{Dynamical evolution process in Scheme II. The figures show the evolution of action combination preferences, and the temporal evolution of $\Delta Q_{s_l}^{M_1,M_2}$. Here, the action combination $(C,C)_D$$(D,D)_C$ indicates that both individuals choose to defect in state C and to cooperate in state D. An exploratory action by one party can disrupt this synchronization, leading to $(C_D$,$D_C)$$(D_C$,$C_D)$, while still maintaining the same action preferences. Thus, $(C,C)_D$$(D,D)_C$ $(C_D$,$D_C)$$(D_C$,$C_D)$ represent the alternation between synchronized and unsynchronized states under the influence of exploratory actions. The action combinations $(C,D)_D$ and $(D,C)_D$ indicate that in state D, one individual cooperates while the other defects. Therefore, $(C,D)_D (D,D)_D$ $(D,C)_D (D,D)_D$ represent this process occurring sequentially and swapping the positions of the two individuals. This can be viewed as an alternating exploitation and punishment process.
Parameters: $\epsilon$ = 0.01, $\alpha$ = 0.1, $\gamma$ = 0.9.
}
\label{fig:Appendix}
\end{figure*}

In Scheme II, the states of both parties directly depend on their respective action information. Therefore, cooperation can be rapidly established only if the random initial conditions fall within the mutual cooperative basin of attraction. If the initial conditions are closer to mutual betrayal, the prerequisite for triggering cooperation is that both parties simultaneously engage in exploratory cooperative behavior. This contrasts with Scheme III, where information can be transmitted through shared states, thereby expediting coordination. This also explains why the convergence time ($t_c$) for high cooperation preference in Scheme III, as depicted in Fig.~\ref{fig:tcTowardsCooperation}(a), is significantly shortened.

To understand the mechanism in Scheme II, we categorize the evolutionary process into three stages based on the characteristics exhibited by the typical time series of $f_c$ shown in Fig.~\ref{fig:TimeSeries_two}.

\begin{enumerate}
\item[1)] Stage i: Mutual betrayal.
\item[2)] Stage ii: Breaking away from mutual betrayal.
\item[3)] Stage iii: Establishing and maintaining mutual cooperation.
\end{enumerate}

Here, $i=\{M_1,M_2\}$ respectively labels the two players, who consider their own action information, the values of $Q_{s_l,a_m}$ and $\Delta Q_{s_l}^{i}$ are labeled in the same way as they are in the text. We initiate the study from initial conditions far from cooperation, i.e., $Q_{C,C}^{M_{1},M_{2}}<Q_{C,D}^{M_{1},M_{2}}$ and $Q_{D,C}^{M_{1},M_{2}}<Q_{D,D}^{M_{1},M_{2}}$, and analyze the mechanism in stages.

\subsection {Stage i --- \emph{Mutual betrayal}}

During this stage, mutual defection $(D, D)_D$ does not yield any payoffs for either party, leading to an increase in $\Delta Q_{D}^{M_1,M_2}$. Intermittent decreases occur due to exploratory cooperation. When $\Delta Q_{D}^{M_1/M_2}> 0$, his/her preference shifts from defection (D) to cooperation (C). However, unilateral cooperation merely strengthens the other player's preference for defection, as the only perceivable change is an increased payoff for maintaining the original action. Therefore, breaking the $(D, D)_D$ preference through a unilateral shift is challenging. As shown in Fig.~\ref{fig:Appendix}(a), $\Delta Q_{D}^{M_0,M_1}$ fluctuates but remains consistently below 0.

\subsection {Stage ii --- \emph{Breaking away from mutual betrayal}}

\textbf{Sub-stage i} -- \emph{Simultaneous cooperative exploration breaks mutual defection, triggers $(C,C)_D$ $\leftrightarrow$ $(D,D)_C$ cyclic state.}

When both individuals simultaneously engage in exploratory cooperative behavior, they achieve positive payoffs $R$, which leads to a continuous increase in $Q_{D,C}^{M_1,M_2}$. This results in $\Delta Q_{D}^{M_1,M_2}>0$ and a reversal in preference [Fig.~\ref{fig:Appendix}(b-2)]. The system then cycles between $(C,C)_D$ $\leftrightarrow$ $(D,D)_C$. Subsequent exploratory behavior disrupts this synchronization, forming the combinations $(C_D,D_C)$ $\leftrightarrow$ $(D_C,C_D)$. Consequently, synchronization and asynchronization alternate [Fig.~\ref{fig:Appendix}(b-1)], with both parties choose to cooperate in state D and defect in state C. However, this action preference combination fails to persist with weak exploration.

\textbf{Sub-stage ii} -- \emph{Alternating exploitation and punishment prepare for establishing cooperation.}

Within the above action preference combination, both individuals' exploratory defection in state D leads to intermittent increases in $Q_{D,D}^{M_1,M_2}$. When $Q_{D,D}^{M_1/M_2} \textgreater Q_{D,C}^{M_1/M_2}$, the system forms the action preference combination $(C,D)_D$ $\leftrightarrow$ $(D_C,D_D)$, with $(D,D)_D$ occurring more frequently. This can be viewed as a process where one party punishes the other, resulting in an increasing trend in $\Delta Q_{C}^{M_1/M_2}$[Fig.~\ref{fig:Appendix}(c)]. When $\Delta Q_{C}^{M_1/M_2}$ \textgreater $0$, the action preference in state C shifts towards cooperation. As indicated by the rectangular dotted boxes, when this process occurs sequentially for both individuals, their preference for defection in state C transitions to cooperation, establishing a $(C,C)_C$ positive feedback loop.

\subsection {Stage iii --- \emph{Establishing and maintaining mutual cooperation}}
In contrast to Scheme III, where individual \emph{Y} continuously exploits individual \emph{M} through asymmetric information, leading to the collapse of cooperation, the case of symmetric information presents a different dynamic. Here, the choice of betrayal by either party directly transitions their respective states to state D. As depicted in Fig.~\ref{fig:Appendix}(d), guided by the Q-table, both parties tend to opt for cooperation in state D, returning to state C and simultaneously enhancing $Q_{D,C}^M$. Consequently, the stability of the cooperative relationship emerges from both parties' propensity to choose cooperation in state D.

\bibliography{aipsamp}

\end{document}